# INTEGRATING SCHEDULABILITY ANALYSIS WITH UML-RT

Q. Gao[*], L.J. Brown[**], L.F. Capretz[**]
[*]Trojan Technology Inc., London, Ontario, N5V 4T7, Canada
[**]University of Western Ontario, London, Ontario, N6A 5B9, Canada
{pgao@trojanuv.com, lbrown@eng.uwo.ca, lcapretz@eng.uwo.ca}

**Abstract:** The use of object oriented techniques and methodologies for the design of real-time control systems appear to be necessary in order to deal with the increasing complexity of such systems. Recently many object-oriented methods have been used for the modeling and design of real-time control systems. We believe that an approach that integrates the advancements in both object modeling and design methods, and real-time scheduling theory is the key to successful use of object oriented technology for real-time software. However, past approaches to integrate the two either restrict the object models, or do not allow sophisticated schedulability analysis techniques. In this paper we show how schedulability analysis can be integrated with object-oriented design; we develop the schedulability and feasibility analysis method for the external messages that may suffer release jitter due to being dispatched by a tick driven scheduler in real-time control system, and we also develop the scheduliability method for sporadic activities, where message arrive sporadically then execute periodically for some bounded time. This method can be used to cope with timing constraints in complex real-time control systems.

**1. Introduction**

There have been many attempts to make use of object-oriented technology for real-time software. Some of them have come from the industrial area [1, 2, 3], whereas others have come from academia [4, 5, 6, 7]. Many of these claims are mostly based on assumption that real-time scheduling theory can be used to perform schedulability analysis. But, traditional real-time scheduling theory results [8, 9, 10, 11] can be directly used only when the object models are restricted to look like the tasking models employed in real-time scheduling theory. In other cases, either the claims are unsupported [2] or based on less sophisticated analysis [4]. Saksena and Karvels [12] provided the first attempt to apply real-time scheduling theory to the object-oriented design by use of the state-of the art in the both fields. In their paper, they show how to integrate traditional scheduliability analysis techniques with object-oriented design models based on the assumptions that the entire external message arrives perfectly on periodic or aperiodic time interval. Martins [13] provided the first attempts to commercially implement scheduling theory for the Unified Modeling Language (UML) by using the technologies in [12], these integrated tools allow issues on timeliness to be addressed much earlier on in the development process.

However, some critical issues regarding real-time control systems are not well addressed by the current approaches, especially because schedulability analysis for real-time control systems has not been effectively incorporated. Although some researchers [12, 13] have addressed this problem by providing code synthesis of scheduling aspects and functionality aspects models, they have mainly focused on the assumptions that all external events arrive perfectly on periodic or aperiodic without release jitter and sporadic effects. In general the real–time control systems do not satisfy these constraints. A message may be delayed by the polling of a tick scheduler, or perhaps awaiting the arrival of another message, and some real-time control systems have messages that behave as so-called sporadically periodic; a message arrives at some time, executes periodically for a bounded number of periods, and then does not re-arrive for a larger time. Examples of such messages are interrupt handlers for burst interrupts or certain monitoring messages in real-time control systems. Until now there is no widely-accepted object-oriented design methodology that deals with these timing constraints for real-time control systems, thus the above analysis methods need to be expanded.

In this paper, we will present an approach to incorporating schedulability analysis in a UML for Real-Time (UML-RT) model-based development process, as an extension of the theoretical work developed by Martins [13]. Using this approach, satisfaction of the end-to-end timing constraints of real-time control systems can be verified and the schedulability analysis results will be used for aspect-oriented code generation in the model transformation and automatic code generation. The rest of the paper is organized as follows. Section 2 introduces schedulability analysis based on RMA. Section 3 describes the feasibility and schedulability analysis methods for real–time control systems with jitter messages and sporadically periodic messages. In 4, we present schedulability results for an example system based on our method. Finally we give with some concluding remarks.

## 2. Schedulability Analysis and Extended Sequence Diagram for UML-RT

Scheduling theory for real-time systems has received a great deal of attention. The first contribution to real-time scheduling theory was made by Liu and Layland [8]. They developed optimal static and dynamic priority scheduling algorithm for hard real-time sets of independent tasks. Since then, significant progress has been made on generalizing and improving the schedulability analysis. The authors developed exact schedulability analysis to determine worst-case timing behavior for tasks with hard real-time constraints in the RMA model considered in the initial work [8], as well as extended models, such as arbitrary deadlines, release jitter, sporadic and periodic tasks [9, 10, 11, 14, 15, 16].

Most of the deterministic schedulability analysis techniques follow the same approach. First, the notion of the critical instant of a task is defined to be an instant at which a request for that task will have the largest response time. Then, the notion of busy period at level '$i$' is defined to be a continuous interval of time during which events of priority '$i$' or higher are being processed [8]. With these concepts, the calculation of the worst-case response time of an action involves the computation of the response time for successive arrivals of the action, starting from a critical instant until the end of the busy period, also the response time of a particular instant of action can be calculated by considering the effects of the blocking factor from lower priority actions and the interference factor from higher or equal priority actions, including previous instances of the same action. If the worst-case response time of the action is less than or equal to its deadline, the action can be said to be schedulable and feasible. Otherwise, the action is not schedulable or feasible.

In our work, we assume that real-time control systems are implemented in a uni-processor single thread environment, and it is made up of a set of transactions, where transaction denotes a single end-to-end computation within the system. Specifically, it refers to the entire causal set of actions executed as a result of the arrival of an external event that originated from an external source. External event sources are typically input devices (such as sensors) that interrupt the CPU-running embedded software. These external events can be periodic or aperiodic, and also have jitter and sporadically periodic characteristics. We express the real-time control system as a collection of transactions that capture all computation in the design model. We also use the term action to capture the processing information associated with an external or internal event. In our model, an action captures this entire run-to-completion processing of an event. The execution of an action may generate internal events that trigger the execution of other actions. Thus, each transaction can be expressed as a collection of actions and events. Each action is a composite action, and composed from primitive sub-actions, these primitive sub-actions include send, call, and return actions [12], which generate internal events through sending messages to other objects.

From UML and UML-RT, we know that the finite state machine behavior models of objects are useful for code-generation; they are not very conducive for reasoning about end-to-end behaviors, or scenarios. UML-RT uses sequence diagrams to model end-to-end system behaviors, or scenarios. However, sequence diagrams are weak in expressing a detailed specification of end-to-end behaviors, which is necessary for schedulability analysis. To express our ideas, we extend the sequence diagram notation to capture detailed end-to-end behaviors.

We use an extended sequence diagram from UML to describe transactions in the system models. In the expanded sequence diagram, we capture the detail of the processing associated with an event. We use the follows notations to represent the different event types.

1. We use " $\rightarrow$ " to represent the asynchronous messages (events).
2. We use " ⟷ " to represent the synchronous messages (events).
3. We use " ⊓⊓ " to represent the periodic messages (events).
4. We use " ⌐⌐ " to represent the aperiodic messages (events).
5. We use " ◆⊓⊓◆ " to represent the sporadically periodic messages (events).
6. We use " ⚡ " to represent the release jitter time of messages (events).

As an illustration, Fig. 1 describes the transaction of automatic gauge control system in a steel mill. The transaction is driven by a timeout message with jitter characteristics. As can be seen, the automatic gauge control object obtains the steel plate thickness from the Thickness Gauge object using a synchronous call action. It then does the control law calculations and generates a position value, which is sent asynchronously to the hydraulic position control object, the hydraulic position control object then sends a command to the hydraulic position actuators adjusting the separation of the rolling cylinders. The sequence diagram for a transaction can easily be expanded to include sub-actions associated with code executed by the real-time execution framework.

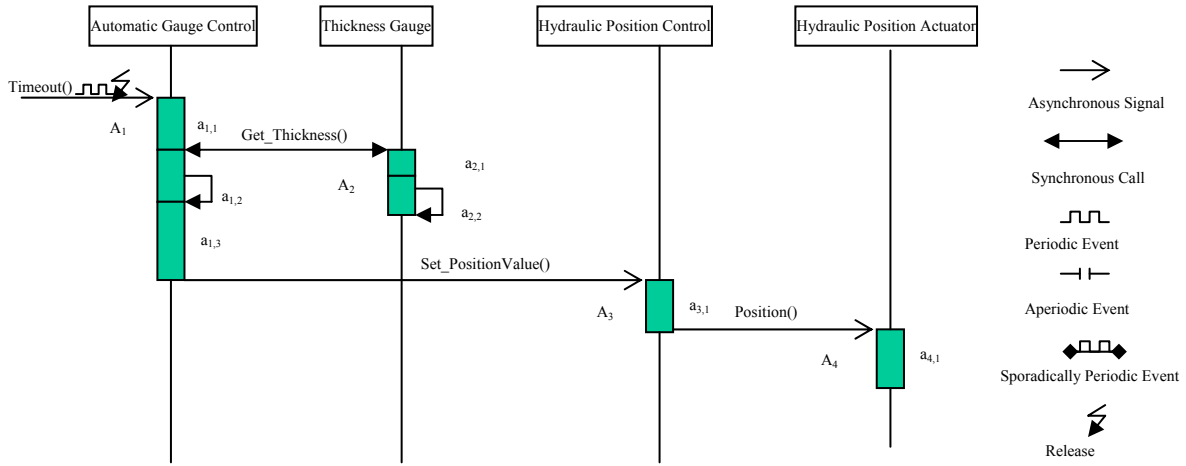

Figure 1: Extended sequence diagram of automatic gauge control system.

In the expanded sequence diagram, we can represent the external events, internal event, actions, and sub-actions. We can also express the external events arrival patterns, such as periodic external event with release jitter, aperiodic event with release jitter, sporadic external event with outer period and inner period. The extended sequence diagram is useful to capture timing constraints such as arrival rates of external events; periodic, aperiodic and sporadically periodic external messages (events); release jitter time of external messages (events); and end-to-end deadlines. This extended sequence diagram has been integrated with a real-time scheduling algorithm to analyze the schedulability and feasibility of control systems. For the purpose of this paper, we are concerned about (1) arrival patterns of the external events, and (2) end-to-end deadlines of actions in the extended sequence diagram. The end-to-end deadlines can be specified on any action in a transaction, which is relative to the arrival of the external event.

**2.1 Notation**

In our paper, we use event and message as synonymous. Let $\varepsilon = \{E_1, E_2, ..., E_n, E_{n+1}, ..., E_N\}$ represent the set of all event-streams in the system, where $E_1, E_2, ..., E_n$ denote external event streams, and the remaining are internal ones. All external events are assumed to be asynchronous, periodic, aperiodic events and sporadic events with release jitter. We use $J_i$ to represent the jitter time of external event $E_i$. $T_i$ and $t_i$ represents the outer period and inner period for sporadically periodic external events $E_i$. If the external event is without sporadic effects, then inner period of such event is equal to its outer period. Each external event stream $E_i$ corresponds to a transaction $\tau_i$.

We also use $A_i$ to represent an action that is associated with each event $E_i$. An action may be decomposed into a sequence of sub-actions $A_i = \{a_{i,1}, a_{i,2}, a_{i,3}, ..., a_{i,n_i}\}$, where each $a_{i,j}$ denotes a primitive action, such as sending a message, calling a message, and returning a message. We use $q$ to represent the instance '$q$' of action $A_i$. Within this model, each action $A_i$ represents the entire "run-to-completion" processing associated with an event $E_i$, and it is characterized as either asynchronously triggered or synchronously triggered, depending on whether the triggering event is asynchronous or synchronous. Each action $A_i$ executes within the context of an active object (capsule) $\tilde{O}(A_i)$, and it is also characterized by a priority ($\pi(A_i)$), which is the same as the priority of its triggering event $E_i$. Each action $A_i$ is also characterized by the computation time ($C(A_i)$) and the deadline ($D(A_i)$). Each sub-action $a_{i,j}$ of $A_i$ is characterized by a computation time $C(a_{i,j})$ (abbreviated as $C_{i,j}$); the computation time of an action is simply the sum of its component sub-actions, i.e., $C(A_i) = \sum_j C_{i,j}$, also, the computation time of any sequential sub-group of sub-actions $a_{i,p}$ to $a_{i,q}$ where $p \leq q$ is $C_{i,p...q} = \sum_{j=p}^{j \leq q} C_{i,j}$. Each event and action is part of a transaction. For the rest of this paper, we use superscript to denote transactions. For example, $A_i^\tau$ represents an action and $E_i^\tau$ represents an event, both of which belong to transaction $\tau$. Adding the superscript for external events $\{E_k : k=1, 2, ..., n\}$ is unnecessary since there is exactly one external event associated with each transaction, i.e., external event $E_k$ belongs to transaction k and would be denoted as $E_k^k$. In this case, the superscript will be omitted.

*2.1.1 Communication Relationships*

We assumed that there are two types of communication relationships between actions, asynchronous and synchronous. We use symbol "→" to denote asynchronous relationship. An asynchronous relationship $A_i \rightarrow A_j$ indicates that action $A_i$ generates an asynchronous signal event $E_j$ (using a send sub-action) that triggers the execution of action $A_j$. Likewise, we use symbol "↔" to denote synchronous relationship. A synchronous relationship $A_i \leftrightarrow A_k$ indicates that action $A_i$ generates a synchronous call event $E_k$ (using a call sub-action) that triggers the execution of action $A_k$. We assume that if the events have a synchronous relationship, the actions have the same priority. We also use a "causes" relationship, and use the symbol $\propto$ for that purpose. Both asynchronous and synchronous relationships are also causes relationships, i.e., $A_i \rightarrow A_j \Rightarrow (A_i \propto A_j)$, and $A_i \Leftrightarrow A_j \Rightarrow (A_i \propto A_j)$, Moreover, the causes relationship is transitive, thus $(A_i \propto A_j) \wedge (A_j \propto A_k) \Rightarrow A_i \propto A_k$. When $A_i \propto A_j$. We say that $A_j$ is a successor of $A_i$ since $A_i$ must execute (at least partially) for $A_j$ to be triggered.

*2.1.2 Synchronous Set*

For the purpose of analysis, we define the term "synchronous set of $A_i$". The synchronous set of $A_i$ is a set of actions that can be built starting from action $A_i$ and adding all actions that are called synchronously from it. The process is repeated recursively until no more actions can be added to the list. We use $\Upsilon(A_i)$ to denote the synchronous set of $A_i$ and $C(\Upsilon(A_i))$ to denote the cumulative execution time of all the actions in this synchronous set. We also call $A_i$ as the root action of this synchronous set.

**2.2 A Case Study**

Fig. 2, for instance, depicts a typical reverse rolling mill in the steel rolling mill. It has a payoff reel, a rolling mill, and a tension reel. A hot coil strip is uncoiled by the payoff reel. The strip is rolled to the specified thickness and coiled by the tension reel. The aim of the rolling process is to reduce the thickness of a strip to a desired thickness gauge. This is done by applying a force to the strip while moving through the roll gap. In order to meet increasing demand for the high precision of strip thickness, a new automatic gauge control system was proposed containing *Roll Gap Control, Roll Speed Control,* and *Roll Eccentricity Compensation.* The *Roll Gap Control System* attempts to adjust the force from the hydraulic cylinder and hence the roll gap, to ensure the output thickness of the rolled strip. The *Roll Speed Control System* automatically adjusts the roll speed according to the mass flow theory and the tension of the steel strip to reduce the influence of thickness fluctuation and satisfy the high quality requirements. The roll *eccentricity compensation system* is applied to adjust the roll gap to accommodate deviations produced as a result of the rolls not being perfectly circular. If the eccentricity compensation is delayed, it can accentuate the errors rather than canceling thus making the strip thickness worse. The eccentricity compensation must be done in the right time or right phase. Even if it is done in the right amplitude, but it is not done at the right time, it can also make the strip thickness worse. All the control systems must guarantee their functional requirements and timing requirements. In order to design such systems, we will use the object–oriented analysis and design methodologies to analysis the functional requirements and timing requirements in such real-time control systems.

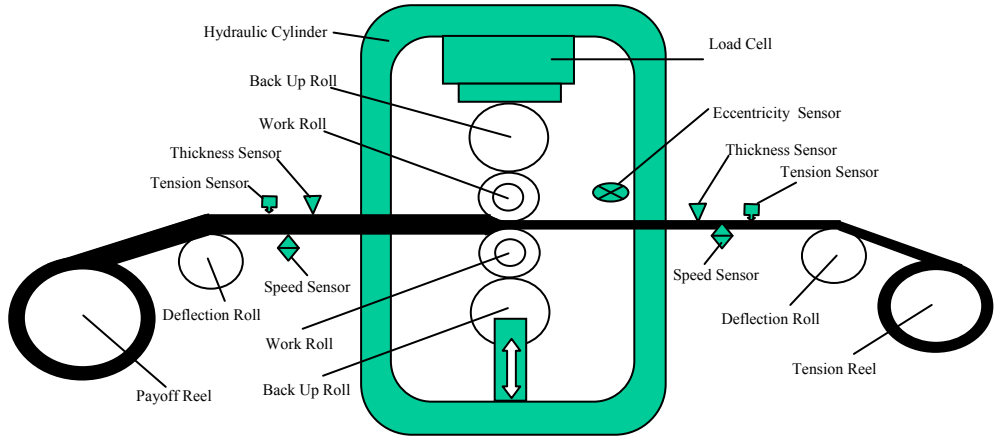

Figure 2. The reverse rolling mill.

*2.2.1 General Description*

Fig. 3 gives the general description of the automatic gauge control system. This system is made up of nine objects, where each object's finite state machine is shown. We can observe that each object has only one "real" state associated with it. We also notice that each object calls its *SpecialInitization* action during initialization, through the system event RTInitSignal, and SpecialDestruction action during system shutdown, through the system event RTDestroySignal. In addition, there are three external events interacting with the system just described above. The first external is thickness setup event. This event is a periodic event with period 60 time unit and 3 time unit release jitter in the system. The second external event is Tension_AGC triggered event, which is an aperiodic event with period 200 time units and 5 time unit release jitter. The third external event is Eccentricity Control Triggered Event; this event is a sporadical event, with outer period 900 time units and inner period 300 time units. The entire external events arrive into the system at time 0.

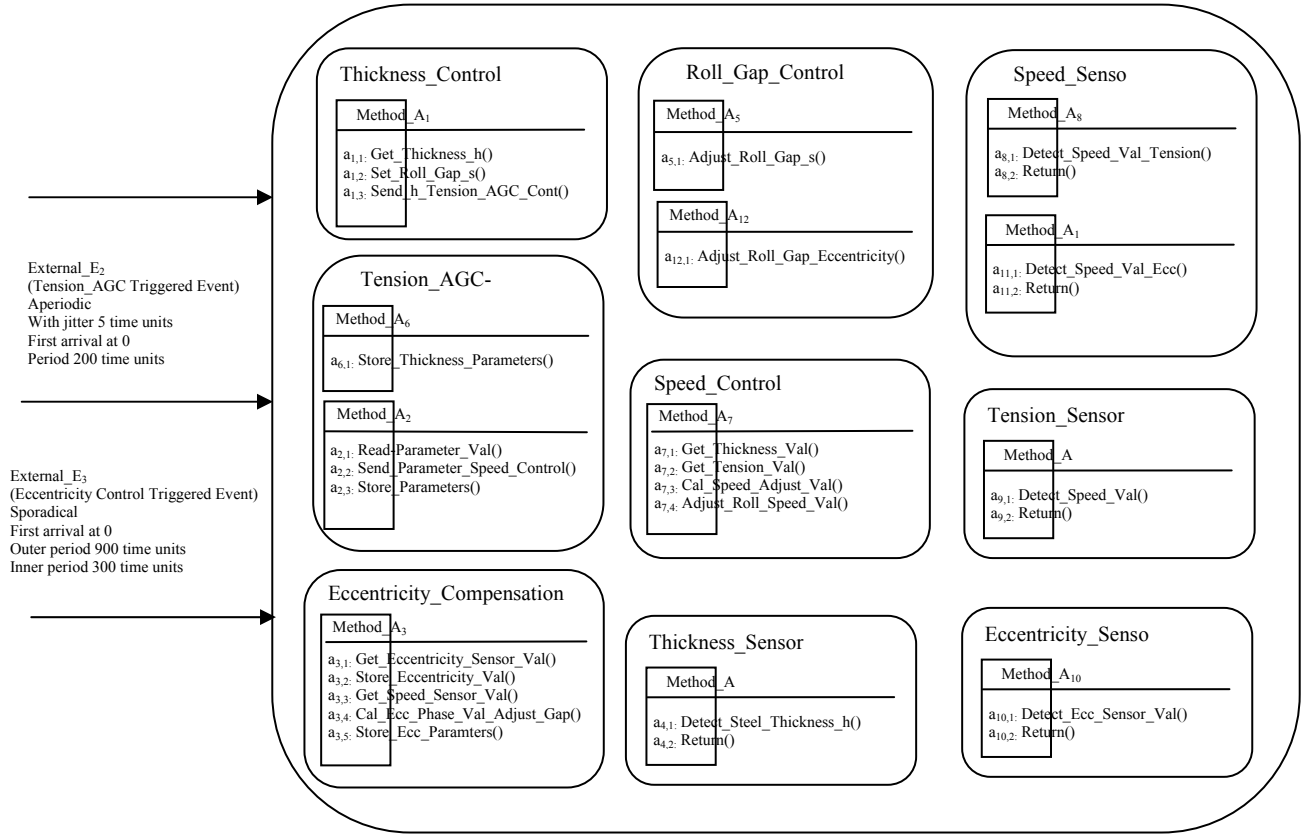

Figure 3. Method description of automatic gauge control system.

*2.2.2 Timing Characteristics of Automatic Gauge Control System*

We have described the automatic gauge control system functional requirements. Now, we will consider the timing characteristics of the system, Table 1 shows the timing characteristics in the automatic gauge control system. All the timing properties can be derived from the real-time control system timing requirements. From Table 1 we can see that events have unique priorities, can arrive at any time, but have variable bounded delay before being placed in a priority-order run-queue. Periodic and aperiodic events are given worst-case inter-arrival time, and sporadically periodic events are given the outer period and inner period. Each event cannot re-arrive sooner than its inner-arrival time; each event may execute a bounded amount of computation, and it is associated with the action, each action is given the worst-case execution time and deadline. This worst-case execution time value is deemed to contain the overhead due to context switching. The cost of pre-emption, within the model, is thus assumed to be zero.

Table 1

Time Characteristic of Automatic Gauge Control System

| Trans $\tau_i$ | Out.P. $T_i$ | Inn.P. $t_i$ | Num. $n_i$ | Jitter $J_i$ | Event(Type) $E_i$ | Action $A_i$ | Priority $\pi(A_i)$ | Deadline $D(A_i)$ | Sub-action $a_{i,j}$ | Comp.Time $C_{i,j}$ | Events Generated $E_i(a_{i,j})$ |
|---|---|---|---|---|---|---|---|---|---|---|---|
| $\tau_1$ | 60 | 60 | 1 | 3 | $E_1$ (External) $E_4$ (call) $E_5$ (Signal) $E_6$ (Call) | $A_1$ $A_4$ $A_5$ $A_6$ | 10 10 10 10 | 60 60 60 60 | $\{a_{1,1},\ a_{1,2},\ a_{1,3}\}$ $\{a_{4,1},\ a_{4,2}\}$ $\{a_{5,1}\}$ $\{a_{6,1}\}$ | $\{5, 1, 1\}$ $\{5, 1\}$ $\{5\}$ $\{3\}$ | $E_4(a_{1,1}),\ E_5(a_{1,2}),\ E_6(a_{1,3})$, --- --- --- |
| $\tau_2$ | 200 | 200 | 1 | 5 | $E_3$ (External) $E_7$ (Signal) $E_8$ (Call) $E_9$ (Call) | $A_2$ $A_7$ $A_8$ $A_9$ | 9 9 9 9 | 125 125 125 125 | $\{a_{2,1},\ a_{2,2},\ a_{2,3}\}$ $\{a_{7,1},\ a_{7,2},\ a_{7,3}, a_{7,4}\}$ $\{a_{8,1},\ a_{8,2}\}$ $\{a_{9,1},\ a_{9,2}\}$ | $\{4,1,5\}$ $\{4,1,5,1\}$ $\{6, 1\}$ $\{8,1\}$ | $E_7(a_{2,2})$ $E_8(a_{7,1}),\ E_9(a_{7,2})$ --- --- |
| $\tau_3$ | 900 | 300 | 3 | | $E_3$ (External) $E_{10}$ (Call) $E_{11}$ (Call) $E_{12}$ (Signal) | $A_3$ $A_{10}$ $A_{11}$ $A_{12}$ | 8 8 8 7 | 250 250 250 250 | $\{a_{3,1}, a_{3,2},\ a_{3,3}, a_{3,4},\ a_{2,5}\}$ $\{a_{10,1},\ a_{10,2}\}$ $\{a_{11,1}\ a_{11,2}\}$ $\{a_{12,1}\}$ | $\{1,3,1,1,4\}$ $\{7,1\}$ $\{6, 1\}$ $\{30\}$ | $E_{10}(a_{3,1}), E_{11}(a_{3,3}), E_{12}(a_{3,4})$ --- --- --- |

*2.2.3 Extended UML-RT for Real-Time Control Systems*

Fig. 4 describes the automatic gauge control system for the No.1 roll stand in the tandem cold rolling mill as discussed. The transaction in the system is driven by different external events. As it can be seen, the Thickness_Control object obtains the steel strip thickness from the Thickness_Sensor object using a synchronous call action. It then does the control law calculations and generates a roll gap value, which is sent asynchronously to the Roll_Gap_Control object, the Roll_Gap_Control object is responsible to adjust the gap of roll in the stand, then using this method to adjust the thickness of steel strip. The extended sequence diagram includes sub-actions associated with code executed by the real-time execution framework. In this extended sequence diagram, we can see the external events, internal event, actions, and sub-actions. We can also express the external event arrival patterns, such as periodic external event with release jitter, aperiodic event with release jitter, sporadic external event with outer period and inner period.

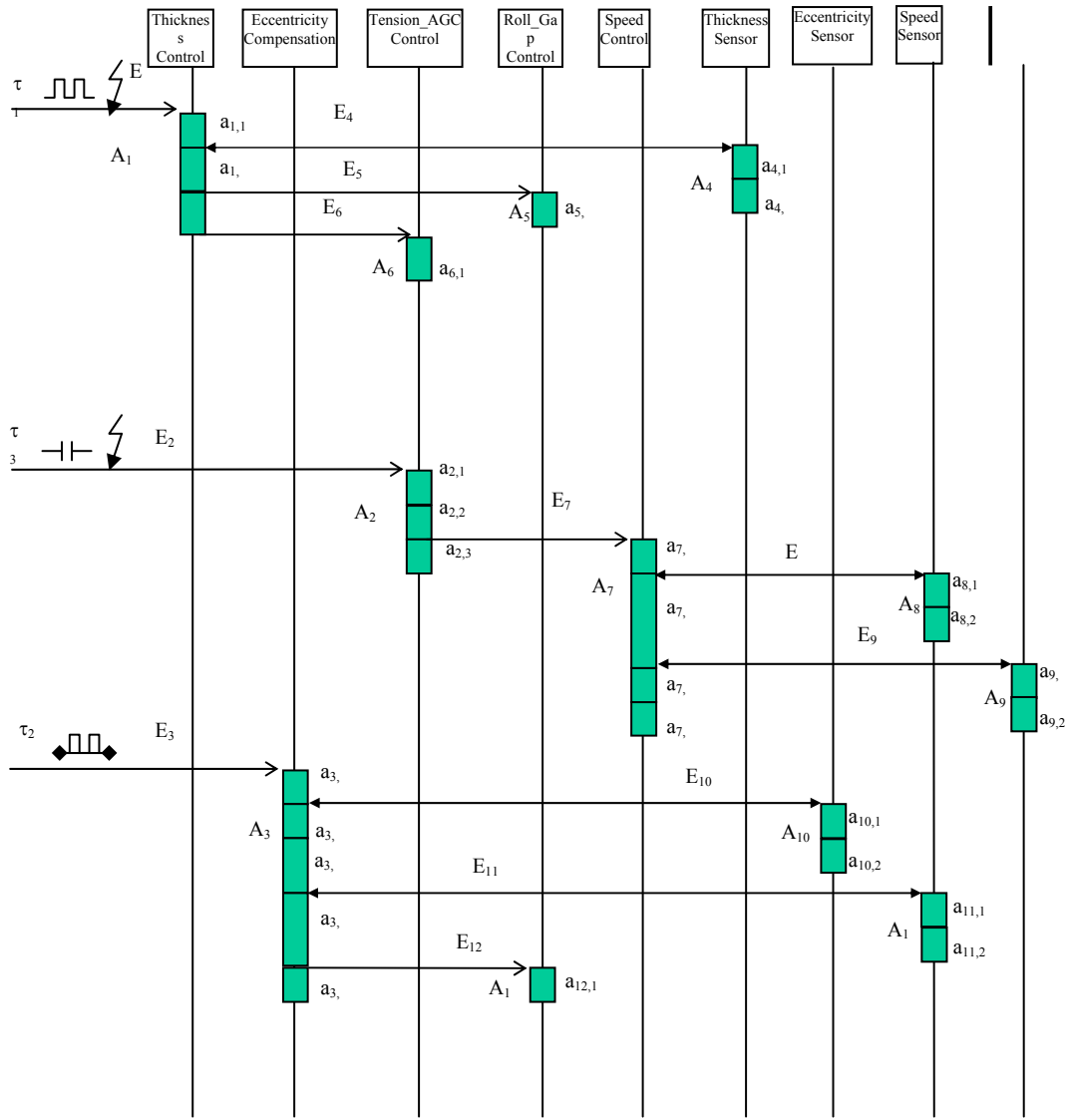

Figure 4. Extended sequence diagram of automatic gauge control system.

## 3. Schedulability and Feasibility Analysis

In our real-time control system model, we assume that only the external events have release jitter, and the internal events do not have jitter, because the internal event arrival is only decided by the action associated with the internal event. For the external events $E_\tau$ which behave as 'sporadically periodic' executing with an inner period ($t_\tau$) and outer period ($T_\tau$). we assume that the 'burst' behavior must finish before the next burst (i.e., $n_\tau t_\tau \leq T_\tau$), where $n_\tau$ is the number of release of external events $E_\tau$ in a burst, and also we assumed that the release jitter ($J_\tau$) of external event $E_\tau$ is the inner release jitter (i.e., each release of external events $E_\tau$ can suffer this jitter).

In our analysis model, we carry out the schedulability and feasibility analysis by calculating the worst-case response time of actions, the worst-case response time of actions $A_i^\tau$ is calculated relative to the arrival of the external event $E_\tau$ that triggers the transaction $\tau$. If the worst-case response time of an action is less than or equal to it's deadline, the action is schedulable, if all the worst-case times of actions in the systems are less than or equal to their deadline; the system is schedulable or feasible. We use the well-known critical instant/busy-period analysis [4, 8, 9, 11] developed for fixed priority scheduling, In our uni-processor single thread implementation environments, a priority inversion occurs if a lower priority event is processed, while a higher priority event is pending. In the same way, a level-*i* busy period is a continuous interval of time during which events of priority "*i*" or higher are being processed.

**3.1 Worst-Case Response Time Analysis**

In the worst-case response time analysis for action $A_i^\tau$, we will compute the response time of the action for successive arrivals of the transaction, staring from a critical instant, until the end of the busy period. We let $S_i^\tau(q)$ denote the worst-case start time for instance '$q$' of action $A_i^\tau$ (i.e., when the instance '$q$' of the action gets the CPU for the first time), starting from the critical instant (time 0). Likewise, $F_i^\tau(q)$ denotes the worst-case finish time, starting from the critical instant (time 0). $A_{rr\tau}(q)$ denotes the arrival time of instance '$q$' of external event $E_\tau$ starting from the critical instant (time 0). According to our system model, we not only consider the busy-period starting at time $J_\tau + qT_\tau$, but also consider the busy-period starting at $J_\tau + q\,t_\tau$ before the release of event $E_\tau$. In order to do that, we define two integers $M_\tau$ and $m_\tau$, where $M_\tau$ is the number of outer periods previously in the window $[0, S_i^\tau(q)]$, and $m_\tau$ is the number of inner periods. $M_\tau$ and $m_\tau$ are given by:

$$M_\tau = \left\lfloor \frac{q-1}{n_\tau} \right\rfloor$$

$$m_\tau = (q-1) - M_\tau m_\tau$$

Where $q$ is an integer, and $q \geq 1$.

The arrival time $A_{rr\tau}(q)$ of instance '$q$' of external event $E_\tau$ can be given as $A_{rr\tau}(q) = M_\tau T_\tau + m_\tau t_\tau$. Based on the traditional scheduling theory for real time systems [8, 9, 10, 11], we can iteratively compute $S_i^\tau(q)$ and $F_i^\tau(q)$ for $q=1,2,3\ldots$ until we reach a $q=m$, such that $F_i^\tau(q) \leq A_{rr\tau}(m+1) - J_\tau$. Then, we let $R(A_i^\tau)$ denote the worst-case response time of action $A_i^\tau$, and it is given by:

$$R(A_i^\tau) = \max_{q \in [1,2,\ldots,m]} \{F_i^\tau(q) + J_\tau - Arr_\tau(q)\}$$

**3.2 Blocking**

According to scheduling theory [8, 12], blocking refers to the effect of lower priority actions on the response time of an action. It may be from any transaction. Let $B(A_i^\tau)$ denote the maximum blocking time of an action $A_i^\tau$. In uni-processor single-thread implementation environments, since scheduling is non-preemptive, priority inversion is limited to one synchronous set of actions with a lower priority root action. This action has started executing just before the transaction containing $A_i^\tau$ arrives. Thus the maximum blocking time of an action is given by:

$$B(A_i^\tau) = \max_{1 \leq k \leq N} \{C(\Upsilon(A_k)) :: \pi(A_i^\tau) \geq \pi(A_k)\}$$

### 3.3 Interference Effects and Busy Period Analysis

We know that the critical instant of an action $A_i^\tau$ occurs when all transaction arrive at the same time (we denote this as time 0), and the root action of the synchronous set of actions that contributes the maximum blocking term $B(A_i^\tau)$. Since actions are executed in a non-preemptive manner, when $A_i^\tau$ starts executing, no other action can interrupt it other than any synchronous calls that $A_i^\tau$ makes. Let early interference function $Early_k^{A_i^\tau(q)}(t)$ denote the interference effect of transaction $k$ prior to $S_i^\tau(q)$, assuming that $S_i^\tau(q) = t$. Then, the value for $S_i^\tau(q)$ is given by the lowest value of $W_i^\tau(q)$, satisfying the following equation:

$$S_i^\tau(q) = \min W_i^\tau(q) :: W_i^\tau(q) = B(A_i^\tau) + \sum_{1 \leq k \leq N} Early_k^{A_i^\tau(q)}(W_i^\tau(q))$$

That is, an action (instance) will start, in the worst case, at a time $W_i^\tau(q)$ if the sum of the blocking and interference effects equals $W_i^\tau(q)$, where $W_i^\tau(q)$ is the first time instant when this become true. Note that the term $W_i^\tau(q)$ occurs on both sides of the equation, this equation can be solved by iteratively refining $W_i^\tau(q)$ using the right side of the equation, starting from an initial lower bound value $B(A_i^\tau)$ in this case, as explained in [8, 12, 15].

Once $S_i^\tau(q)$ is known, we can compute $F_i^\tau(q)$. Therefore, $F_i^\tau(q)$ can be calculated as follow:

$$F_i^\tau(q) = S_i^\tau(q) + \mathbf{C}(\Upsilon(\mathbf{A}_i^\tau))$$

Where $\mathbf{C}(\Upsilon(\mathbf{A}_i^\tau))$ is the cumulative execution time of all the actions in this synchronous set of $A_i^\tau$.

### 3.4 Early Interference Function

The early interference function depends on whether we are considering interference from a different transaction, i.e., $k \neq \tau$, or from the same transaction, i.e., $k = \tau$.

*3.4.1 Early Interference effects from Different Transactions*

In this case, we consider the arrival of transactions where $k \neq \tau$ in the interval $[0, W_i^\tau(q)]$. We have to consider the computation times of all higher or equal priority actions making up transaction $k$. Again, any synchronous call made recursively from the resulting actions will be considered, because of our earlier assumption that the priority of a synchronously triggered action is the same as that of the caller action. Note that we have to take the closed interval, because if a higher action becomes enabled at time $W_i^\tau(q)$, then $A_i^\tau(q)$ cannot begin executing. Now consider the computation occurring in the window $[0, W_i^\tau(q)]$ from higher priority sporadically periodic event $E_k$ with release jitter $J_k$. If the window is larger than the number of 'bursts' of $E_k$ then the computation time from each burst amount is $n_k C(A_k)$. For the partial 'burst' starting in the window, we can treat $E_k$ as a simple periodic event executing with period $t_k$ over the remaining part of the window. We let $F_K$ represent the whole number of event $E_k$ 'bursts' starting and finishing in the window, and it is given as follow:

$$\mathbf{F}_k = \left\lfloor \frac{J_k + W_i^\tau(q)}{T_k} \right\rfloor$$

The remaining part of the window $[0, W_i^\tau(q)]$ is the length $J_K + W_i^\tau(q) - F_k T_k$. Hence a bound on the number of events $E_k$ in this remaining time is $F_{kr}$, and it is given by:

$$\mathbf{F}_{kr} = \left\lfloor \frac{J_K + W_i^\tau(q) - F_K T_K}{t_K} \right\rfloor + 1$$

Another bound on the number of events $E_k$ in this remaining time is $n_k$, since a burst can consist of at most $n_k$ invocations of event $E_k$. Therefore the least upper bound number $F_{kr\min}$ can be given by:

$$\mathbf{F}_{kr\min} = \min(n_k, \mathbf{F}_{kr})$$

So the total interference of action $A_i^\tau$ from different transaction $k$ is given as:

$$\mathbf{Early}_{k \neq \tau}^{A_i^\tau(q)}(W_i^\tau(q)) = (\mathbf{F}_{kr\min} + \mathbf{F}_K n_k) \bullet \sum_l (c(A_l^k) :: \pi(A_l^k) \geq \pi(A_i^\tau))$$

*3.4.2 Early Interference effects from the Same Transaction*

In this case, we consider the arrival of transactions where $k = \tau$ in the interval $[0, W_i^\tau(q)]$. It is important to distinguish between previous instances, i.e., 1,2, …, q-1 of the transaction, and all other instances after that. Accordingly, we can write:

$$\mathbf{Early}_\tau^{A_i^\tau(q)}(W_i^\tau(q)) = \mathbf{Early}_{\tau^-}^{A_i^\tau(q)}(W_i^\tau(q)) + \mathbf{Early}_{\tau^+}^{A_i^\tau(q)}(W_i^\tau(q))$$

Where the $\mathbf{Early}_{\tau^-}^{A_i^\tau(q)}(W_i^\tau(q))$ is the interference effects from the past instances (1,2,…, q-1) and $\mathbf{Early}_{\tau^+}^{A_i^\tau(q)}(W_i^\tau(q))$ is the interference effects of all other instances $q, q+1,…$ that may have arrived in $[0, S_i^\tau(q)]$.

The past instances of the transaction have similar effects as other transactions, since any higher or equal priority actions of the transaction must execute prior to $A_i^\tau(q)$. Thus the Early$_{\tau^-}^{A_i^\tau(q)}(W_i^\tau(q)$ can be given as:

$$\textbf{Early}_{\tau^-}^{A_i^\tau(q)}(W_i^\tau(q)) = (M_\tau n_\tau + m_\tau) \bullet \sum_l (C(A_l^\tau) :: \pi(A_l^\tau) \geq \pi(A_i^\tau))$$

The interference effect of instance $q$ onwards must not count the effect of any action $A_l^\tau$, if $A_i^\tau \propto A_l^\tau$, since if $A_i^\tau(q)$ has not executed, any action that is caused by it could not have executed either. Furthermore, we assume that multiple instances of the same action execute in order and thus, this is true for instance $q+1$ onward as well.

If the action $A_i^\tau$ is asynchronously triggered, the Early$_{\tau^+}^{A_i^\tau(q)}(W_i^\tau(q))$ is given by the following equations. Let $F_\tau$ represent the whole number of events $E_\tau$ 'bursts' starting and finishing in the window $[0, W_i^\tau(q)]$ and is given by:

$$F_\tau = \left\lfloor \frac{W_i^\tau(q)}{T_\tau} \right\rfloor$$

The remaining part of the window $[0, W_i^\tau(q)]$ is the length $W_i^\tau(q) - F_\tau T_\tau$, hence a bound on the number of events $E_\tau$ in this remaining time is $F_{\tau r}$, and it is given by:

$$F_{\tau r} = \left\lfloor \frac{W_i^\tau(q) - F_\tau T_\tau}{t_\tau} \right\rfloor + 1$$

Another bound on the number of events $E_\tau$ in this remaining time is $n_\tau$, since a burst can consist of at most $n_\tau$ invocations of event $E_\tau$. Therefore the least upper bound number $F_{\tau r \min}$ can be given by:

$$\textbf{F}_{\tau r \min} = \min(n_\tau, \textbf{F}_{\tau r})$$

So the Early$_{\tau^+}^{A_i^\tau(q)}(W_i^\tau(q))$ is given by:

$$\mathbf{Early}_{\tau^+}^{A_i^\tau(q)}(\mathbf{W}_i^\tau(q)) = \{(\mathbf{F}_{\tau r \min} + \mathbf{F}_\tau \mathbf{n}_\tau) -$$

$$(M_\tau n_\tau + m_\tau)\} \bullet (\sum_l (C(A_l^\tau) :: \neg(A_i^\tau \propto A_l^\tau) \wedge \pi(A_l^\tau) \geq \pi(A_i^\tau))$$

According to the above analysis, for the asynchronously triggered action $A_i^\tau$, we can find start times $S_i^\tau(q)$ as follows:

$$S_i^\tau(q) = \min \mathbf{W}_i^\tau(q) ::$$

$$\mathbf{W}_i^\tau(q) = \mathbf{B}(\mathbf{A}_i^\tau) + \sum_{1 \leq k \leq N} Early_k^{A_i^\tau(q)}(\mathbf{W}_i^\tau(q))$$

$$= \mathbf{B}(\mathbf{A}_i^\tau) + \sum_{\substack{k \neq \tau \\ 1 \leq k \leq N}} (F_{kr \min} + F_k n_k) \sum_l (c(A_l^K) :: \pi(\mathbf{A}_l^k) \geq \pi(\mathbf{A}_i^\tau))$$

$$+ (M_\tau n_\tau + m_\tau) \bullet \sum_l (C(A_l^\tau) :: \pi(\mathbf{A}_l^\tau) \geq \pi(\mathbf{A}_i^\tau))$$

$$+ \{(\mathbf{F}_{\tau r \min} + \mathbf{F}_\tau \mathbf{n}_\tau) - (M_\tau n_\tau + m_\tau)\} \bullet (\sum_l (C(A_l^\tau) :: \neg(A_i^\tau \propto A_l^\tau) \wedge \pi(A_l^\tau) \geq \pi(A_i^\tau))$$

If the action $A_i^\tau$ is synchronously triggered, the above worst staring time $S_i^\tau(q)$ for the asynchronously triggered action $A_i^\tau$ may be improved. Consider a synchronously triggered action $A_i^\tau$, let $A_g^\tau$ be the asynchronously triggered action, such that $A_i^\tau$ belongs to $\Upsilon(A_g^\tau)$, i.e., the synchronous-set of $A_g^\tau$. Then we have a chain of actions, starting from $A_g^\tau$ to $A_i^\tau$ that only execute partially in this interval, and are blocked waiting for $A_i^\tau$ to execute. Note that there must be exactly one such action $A_g^\tau$, so there is no ambiguity. This changes the interference for instances $q$, $q+1$, ... of transaction τ. For instance $q$, only a part of the synchronous set $\Upsilon(A_g^\tau)$ has executed, and this should be reflected in the equation. Rather than extending the notation to explicitly define this subset, we denote the sub-action producing the action $A_i^\tau$ as $a_{g,h}^\tau$, and the computation time associated with this sub-action as $C(sub(\gamma(a_{g,1...h}^{\tau})))$. For instances $q+1$ onwards, none of the actions in the synchronous set $\Upsilon(A_g^{\tau})$ can cause interference, since their previous instance ($q$) is blocked. The blocking term, interference from other transaction, and interference from previous instances (0,1,2, ...,$q-1$) of the same transaction remain the same, because we assumed that $\pi(A_g^{\tau}) = \pi(A_i^{\tau})$. Based on the above analysis, the worst starting time $S_i^{\tau}(q)$ for the synchronously triggered action $A_i^{\tau}$ is given as follows:

$$S_i^{\tau}(q) = \min \mathbf{W}_i^{\tau}(q) ::$$

$$\mathbf{W}_i^{\tau}(q) = \mathbf{B}(A_g^{\tau}) + \sum_{1 \leq k \leq N} Early_k^{A_i^{\tau}(q)}(\mathbf{W}_i^{\tau}(q))$$

$$= \mathbf{B}(A_g^{\tau}) + \sum_{\substack{k \neq \tau \\ 1 \leq k \leq N}} (F_{kr\min} + F_k n_k) \sum_l (c(A_l^K) :: \pi(A_l^k) \geq \pi(A_g^{\tau}))$$

$$+ (M_{\tau} n_{\tau} + m_{\tau}) \cdot \sum_l (C(A_l^{\tau}) :: \pi(A_l^{\tau}) \geq \pi(A_g^{\tau}))$$

$$+ C(sub(\gamma(a_{g,1...h}^{\tau}))) + \sum_l (C(A_l^{\tau}) :: \neg(A_g^{\tau} \propto A_l^{\tau}) \wedge \pi(A_l^{\tau}) \geq \pi(A_g^{\tau})$$

$$+ \{(F_{\tau r\min} + F_{\tau} n_{\tau}) - (M_{\tau} n_{\tau} + m_{\tau}) - 1\} \cdot (\sum_l (C(A_l^{\tau}) :: \neg(A_i^{\tau} \propto A_l^{\tau}) \wedge \pi(A_l^{\tau}) \geq \pi(A_g^{\tau})$$

**4. Schedulability Analysis**

From the above equations, we can calculate the value of $S_i^{\tau}(q)$. Once the value of $S_i^{\tau}(q)$ is obtained from the above equations, we can iteratively compute $S_i^{\tau}(q)$ and $F_i^{\tau}(q)$ for $q=1,2,3$ ..., until we reach a $q=m$, such that $F_i^{\tau}(q) \leq A_{rr\tau}(m+1) - J_{\tau}$. Then, the worst-case response time of action $A_i^{\tau}$ is given by:

$$R(A_i^{\tau}) = \max_{q \in [1,2,...,m]} \{F_i^{\tau}(q) + J_{\tau} - Arr_{\tau}(q)\}$$

If the worst-case response time $R(A_i^r)$ is less than or equal to it's deadline $D(A_i^r)$, then the action $A_i^r$ implementation is feasible. If the worst-case response time $R(A_i^r)$ is larger than the deadline $D(A_i^r)$, then the action $A_i^r$ implementation is not feasible. If all the action worst-case response times in the real-time control system are less than or equal to their deadlines, we can say that the systems implementation is feasible.

**4.1 Schedulability Analysis for our Case Study**

Now, let us revisit our automatic gauge control system and apply the above scheduling analysis method to analyze the system schedulability. Table 2 shows the worst-case response time of each action found by this analysis method. From that table, we can see that all the worst- case response times of actions in the system are less than their deadline constraint, so we can say that the system is schedulable and feasible. From the results, we can also see that the worst-case response time of all actions is large due to the action $A_{12}$ which has large computation time and the lowest priority in the system. Since in our system model, the implementation is in uni-processor single thread environment, it causes blocking for all other actions.

Based on the results, we can see that the effect of the lowest priorities of action $A_{12}$ is also reflected in its larger worst-case response time because of the greater interference. For non-preemptive scheduling in our uni-processor single thread environment, the worst-case response time of the lowest priority action $A_{12}$ is relatively large. Once the action starts executing, it executes as if its priority is raised to the highest priority in the system. From the results, we can also see that the worst-case response time of action $A_3$ has the largest worst-case response time. This is because that it is affected by the higher priority interference and lower priority blocking, it has two synchronously call sub-actions and it must wait for the recipient action to finish their execution.

Table 2

Worst-Case Response Time of Automatic Gauge Control System

| Transaction | Action | Priority | Deadline | Worst-Case Response Time |
|---|---|---|---|---|
| $\tau_1$ | $A_1$ | 10 | 60 | 46 |
| | $A_4$ | 10 | 60 | 39 |
| | $A_5$ | 10 | 60 | 38 |
| | $A_6$ | 10 | 60 | 36 |
| $\tau_2$ | $A_2$ | 9 | 125 | 114 |
| | $A_7$ | 9 | 125 | 114 |
| | $A_8$ | 9 | 125 | 94 |
| | $A_9$ | 9 | 125 | 96 |
| $\tau_3$ | $A_3$ | 8 | 250 | 122 |
| | $A_{10}$ | 8 | 250 | 117 |
| | $A_{11}$ | 8 | 250 | 116 |
| | $A_{12}$ | 7 | 250 | 119 |

If we change the priority of action 7 from 9 to 8 in the automatic gauge control system, i.e., changing the priority of action 7 from higher priority to lower priority, we get the worst-case response time results as shown in Table 3. We can see in Table 3 that the worst case response time of action 7 changes to 139 seconds from 114 seconds. Even though all the worst-case response time of other actions are less than their deadline constraint, we cannot say that the system is feasible because the worst-case response time of action 7 is larger than its deadline.

From the above analysis results, we can see that our extended schedulability analysis method can be used to analyze the schedulability and feasibility of real-time control systems with release jitter and sporadic effects. Using this method, a designer can quickly evaluate the impact of various implementation decisions on schedulability. In conjunction with automatic code generation, this can greatly reduce the development of real-time control system software.

Table 3

Worst-Case Response Time of an Unfeasible Automatic Gauge Control System

| Transaction | Action | Priority | Deadline (Sec.) | Worst Case Response Time (Sec.) |
|---|---|---|---|---|
| $\tau_1$ | $A_1$ | 10 | 60 | 46 |
| | $A_4$ | 10 | 60 | 39 |
| | $A_5$ | 10 | 60 | 38 |
| | $A_6$ | 10 | 60 | 36 |
| $\tau_2$ | $A_2$ | 9 | 125 | 66 |
| | $A_7$ | 8 | 125 | 139 |
| | $A_8$ | 8 | 125 | 119 |
| | $A_9$ | 8 | 125 | 121 |
| $\tau_3$ | $A_3$ | 8 | 150 | 122 |
| | $A_{10}$ | 8 | 150 | 117 |
| | $A_{11}$ | 8 | 150 | 116 |
| | $A_{12}$ | 7 | 150 | 119 |

**5. Final Remarks**

Software design has become more and more important within the real-time control system design process since functionality implementation gradually migrated from hardware to software. Consequently, several commercial tools have become available that provide an integrated development environment for real-time control systems with object-oriented techniques to facilitate the design phase. However, these tools lack the 'real-time" support required by many of these systems, especially those with stringent timing constraints.

As a result, we proposed a methodology for the integration of schedilability analysis techniques within UML-RT techniques to support the timing requirements in real-time control system design process. The main contribution of our paper is in the development of the worst-case response time analysis for object-oriented design models in which the external events suffer release jitter and have sporadically periodic characteristics. We also extended UML sequence diagrams to visually describe the timing properties for real-time control systems. The results developed are also generally applicable to any modeling language using active objects, and explicit communication between objects through message passing. This method can be used to cope with timing constraints in realistic and complex real time control systems. Using this method, a designer can quickly evaluate the impact of various implementation decisions on schedulability. In conjunction with automatic code-generation, we believe that this will greatly streamline the design and development of real-time control system software.

**Authors' Biographies:**

*Q. Gao* received his B.Eng. (Electrical and Computer Engineering) from Anshan University of Science and Technology, China, and M.E.Sc. from the University of Western Ontario, Canada, in 1986 and 2003 respectively. Mr. Gao worked in the Shenyang Institute of Science in a Chinese project to develop an autonomous underwater vehicle and remotely operated vehicles. Currently, he is a controls engineer with Trojan Technologies Inc, in London, Ontario, where he designs real-time control systems for ultraviolet (UV) water and wastewater treatment series products. His has research interest in software for real-time control systems in the area of industrial process control and automation.

*L.J. Brown* received the B.E.Sc. degree in electrical engineering from the University of Waterloo, Waterloo, ON, Canada, in 1988 and the M.S. and Ph.D. degrees in electrical engineering from the University of Illinois, Urbana-Champaign, in 1991 and 1996, respectively. He was with E. I. DuPont de Nemours, Newark, DE, from 1996 to 1999 and joined the Faculty of Engineering, University of Western Ontario, London, ON, in 1999. His research interests include adaptive control, periodic signals, manufacturing control, welding control and biological control systems.

*L.F. Capretz* has over 20 years of experience in the software engineering field as a practitioner, manager and educator. Before joining the University of Western Ontario, in Canada, he worked at both technical and managerial levels, taught and carried out research on the engineering of software in Brazil, Argentina, England and Japan since 1981. He was the Director of Informatics and Coordinator of the computer science program in two universities (UMC and COC) in the State of Sao Paulo/Brazil. He has authored and co-authored over 50 peer-reviewed research papers on software engineering in leading international journals and conference proceedings, and co-authored the book, *Object-Oriented Software: Design an Maintenance*, published by World Scientific. His current research interests are software engineering (SE), human factors in SE, software product lines, and software engineering education. Dr. Capretz received his Ph.D. from the University of Newcastle upon Tyne (U.K.), Master's from the National Institute for Space Research (INPE-Brazil), and B.Sc. from UNICAMP (Brazil). He is a senior member of IEEE.